\documentclass[11pt,nofootinbib,showpacs]{revtex4}
\usepackage{amsfonts,amssymb,amsmath,graphicx,multirow}
\def\dac{\displaystyle\frac}

\def\[{\left[}
\def\]{\right]}
\def\({\left(}
\def\){\right)}

\newcommand{\const}{\mathop{\rm const}\nolimits}
\newcommand{\Tr}{\mathop{\rm Tr}\nolimits}
\newcommand{\kK}{\kappa K}
\begin{document}

\baselineskip7mm
\title{Dynamics of gravitating hadron matter in Bianchi-IX cosmological model}

\author{Sergey A. Pavluchenko}
\affiliation{Programa de P\'os-Gradua\c{c}\~ao em F\'isica, Universidade Federal do Maranh\~ao (UFMA), 65085-580, S\~ao Lu\'is, Maranh\~ao, Brazil}

\begin{abstract}
We perform an analysis of the Einstein-Skyrme cosmological model in Bianchi-IX background. We analytically describe asymptotic regimes and semi-analytically -- generic regimes. It appears that depending on
the product of Newtonian constant $\kappa$ with Skyrme coupling $K$, in absence of the cosmological term there are three possible regimes -- recollapse with $\kK < 2$ and two power-law regimes --
$\propto t^{1/2}$ for $\kK=2$ and $\propto t$ for $\kK > 2$. In presence of the positive cosmological term, power-law regimes turn to the exponential (de Sitter) ones while recollapse regime turn to exponential
if the value for $\Lambda$-term is sufficiently large, otherwise the regime remains recollapse. Negative cosmological term leads to the recollapse regardless of $\kK$.
All nonsingular regimes have the squashing coefficient $a(t) \to 1$ at late times, which is associated with restoring symmetry dynamics. Also all nonsingular regimes appear to be linearly stable --
exponential solutions always while power-law for an open region of the initial conditions.
\end{abstract}



\maketitle

\section{Introduction}

One of the most important nonlinear field theories is the sigma model, with its applications covering many aspects of quantum physics (see e.g. \cite{Manton1} for review), but within this model
it is impossible to build static soliton solutions in 3+1 dimensions. To overcome this Skyrme introduced~\cite{Skyrme} term which allows static soliton solutions with finite energy, called {\it Skyrmions}
(see also \cite{Manton1, ped2} for review), to exist. It appears that excitations around Skyrme solutions may represent Fermionic degrees of freedom, suitable to describe baryons (see~\cite{ferm1} for
detailed calculations and~\cite{ferm2, ferm3, ferm4, ferm5} for examples). Winding number of Skyrmions is identified with the baryon number in particle physics~\cite{wind}.
Apart from particle and nuclear physics, Skyrme theory is relevant to astrophysics~\cite{astro},
Bose-Einstein condensates~\cite{EBc}, nematic liquids~\cite{nematic}, magnetic structures~\cite{magnetic} and condensed matter physics~\cite{cond}. Also, Skyrme theory naturally appears in AdS/CFT
context~\cite{adscft}.

Due to highly nonlinear character of sigma and Skyrme models, it is very difficult to build exact solution in both of them. So, to make field equations more tractable, one usually adopts certain {\it ansatz}.
For Skyrme model one of the best known and mostly used is hedgehog {\it ansatz} for spherically symmetric systems, which reduces field equations to a single scalar equation. It worth mentioning that recently
this {\it ansatz} was generalized~\cite{CH} for non-spherically-symmetric cases.

Use of hedgehog {\it ansatz} allows study of self-gravitating Skyrme models. In particular, it was demonstrated the potential presence of Skyrme hair for spherically-symmetric black-hole
configurations~\cite{hair}. This is the first genuine counterexample to ``no-hair'' conjecture which appears to be stable~\cite{h1}; its particle-like~\cite{h2} counterparts and dynamical
configurations~\cite{h3} have been studied numerically. After that, more realistic spherically- and axially-symmetric black-hole and regular configurations were studied~\cite{recent}.

Apart from spherically-symmetric configurations, of particular interest are cosmologically-type solutions. Generalized hedgehog {\it ansatz} makes it possible to write down simplified field equations for
non-spherically-symmetric configurations which we used to perform analysis of Bianchi-I and Kantowski-Sachs models for Einstein-Skyrme cosmology with $\Lambda$-term~\cite{we14} (particular subcase was studied
in~\cite{another}). The paper~\cite{fabr15} was a logical continuation of them, as the particular solution of the Bianchi-IX cosmological model was described. The analysis suggests that, based on the static
counterpart of this model, the construction of exact multi-Skyrmion configurations composed
by elementary spherically symmetric Skyrmions with non-trivial winding number in four-dimensions is possible~\cite{rest} (see also~\cite{sun} for possible generalization to higher $SU(N)$ models).

In this paper we are going to consider full Bianchi-IX cosmological model in Einstein-Skyrme system. Our study is motivated from both field theory and cosmological point of view. Indeed, this is one of few
(if not the only) systems where one can study analytically dynamical and cosmological consequences of the conserved topological charge, which in this particular case is associated with the baryon number.
From the cosmological point of view, Bianchi-IX model is well-known and well-studied in cosmology -- for instance, for the proof of inevitability of the physical singularity through oscillatory approach to
it~\cite{belinski}. So that, if we consider Bianchi-IX model, the results could be translated and compared with the counterparts from our physical Universe.

The structure of the manuscript is as follows: first we review Einstein-Skyrme system and derive basic equations, then we study asymptotic case both with and without $\Lambda$-term. After that we study
general case, address linear stability of the obtained solutions and finally discuss and summarize the results.

\section{Equations of motion}

The Skyrme action can be constructed in the following way: let be $U$ a
$SU(2)$ valued scalar field. We can the define the quantities:
\begin{equation*}
A_{\mu}^i t_i \equiv A_{\mu}=U^{-1}\nabla_{\mu}U,
\end{equation*}
\begin{equation*}
F_{\mu \nu}=[A_{\mu},A_{\nu}].
\end{equation*}
Here the Latin indices correspond to the group indices and the generators $t_i$ of $SU(2)$ are related to the Pauli matrices by $t_i =-i \sigma_i$.
The Skyrme action is then defined as
\begin{equation}
S_{Skyrme}=\frac{K}{2}\int d^4 x\sqrt{-g}\mathrm{Tr}\left( \frac{1}{2}A_{\mu}A^{\mu}+ \frac{\lambda}{16}F_{\mu \nu}F^{\mu \nu} \right). \label{s1}
\end{equation}
The case when $\lambda =0$ is called non linear Sigma Model and the term which multiplies $\lambda$ is called the Skyrme term. The total action for a self gravitating Skyrme field reads
\begin{equation}
S=\int d^4x \sqrt{-g}\dac{R-2\Lambda}{2\kappa}+S_{Skyrme},  \label{s2}
\end{equation}
where $\kappa$ is the gravitational constant, $R$ is the Ricci scalar and $\Lambda$ is the cosmological constant.
Skyrme field equation reads
\begin{equation}
\nabla^{\mu}A_{\mu}+\frac{\lambda}{4}\nabla^{\mu}[A^{\nu},F_{\mu\nu}]=0. \label{skyrme}
\end{equation}

The topological charge of the Skyrme model is
\begin{equation}
w = - \dac{1}{24\pi^2} \int\limits_{t=\const} \Tr \[ \epsilon^{ijk} A_i A_j A_k \],  \label{top}
\end{equation}
and physically it represents the baryonic charge.

The $SU(2)$ valued scalar field can be parameterized in a standard way
\begin{equation*}
U=\mathbf{I} Y^0 + Y^it_i \; \; \; ; \; \; \; U^{-1}= \mathbf{I} Y^0 - Y^it_i, \label{Ufield}
\end{equation*}
with $Y^0=Y^0(x^{\mu})$ and $Y^i=Y^i(x^{\mu})$ must satisfy $(Y^0)^2+Y_iY^i=1$.
The most famous and most studied {\it ansatz} for searching solutions to the (non-self gravitating) Skyrme theory is so called ``hedgehog'' which is obtained by choosing
\begin{equation*}
 Y^0 =\cos (\alpha ) \; \; \; ; \; \; \;  Y^i= n^i\sin (\alpha ), \label{hedgehog}
\end{equation*}
where $\alpha$ is a radial profile function and $n^i$ is a normal radial vector
\begin{equation*}
n^1=\sin \Theta \cos \Phi \; \; \; ; \; \; \; n^2 = \sin \Theta \sin \Phi \; \; \; ; \; \; \; n^3=\cos \Theta. \label{normalvector}
\end{equation*}

As mentioned, we work with Bianchi-IX metric

\begin{equation}
ds^2=-dt^2+\frac{\rho^2 (t)}{4}\left[ a^2 (t)(d\gamma +\cos \theta d\varphi )^{2}+d\theta ^{2}+\sin ^{2}\theta d\varphi ^{2}\right] , \label{BIX}
\end{equation}
where $\rho (t)$ is a global scale factor and $a(t)$ is a squashing coefficient. One can check that (see also~\cite{fabr15}), with unit baryonic charge $w=+1$ (\ref{top}), the configuration
\begin{equation}
\Phi =\frac{\gamma +\varphi }{2},\ \tan \Theta =\frac{\cot \left( \frac{\theta }{2}\right) }{\cos \left( \frac{\gamma -\varphi }{2}\right) },\ \tan \alpha =\frac{\sqrt{1+\tan ^{2}\Theta }}{\tan \left( \frac{\gamma -\varphi }{2}\right) }  \label{conf}
\end{equation}
identically satisfies the Skyrme field equations (\ref{skyrme}) on any background metric of the form (\ref{BIX}). Now substituting metric (\ref{BIX}) and configuration (\ref{conf}) into action (\ref{s1}) and (\ref{s2})
as well as to the hedgehog {\it ansatz}, one can derive equations of motion in the following form (see also~\cite{fabr15}):

\begin{equation}
\begin{array}{l}
2a\rho ^2 (2\rho \dot{a}+3a\dot{\rho})\dot{\rho}-2a^2 \rho^2 (\Lambda \rho^2 +a^2 -4)-\kappa K[(2\rho^2 +\lambda) a^2 +\rho^2 +2\lambda ] =0,  \\
2a^2 \rho^2 (2\rho \ddot{\rho}+\dot{\rho}^2)-2a^2 \rho^2 (\Lambda \rho^2 +3a^2 -4)-\kappa K[(2\rho^2 +\lambda)a^2 -\rho^2 -2\lambda] =0,   \\
a\rho^3 (\rho \ddot{a}+3\dot{\rho}\dot{a})+(a^2 -1)[\kappa K(\lambda +\rho^2)+4a^2\rho^2] =0.
\end{array} \label{full}
\end{equation}

\section{Asymptotic $a(t) \equiv 1$ case}

We start from equations for the special case $a(t) \equiv 1$ after substituting it into~(\ref{full}):

\begin{equation}
\begin{array}{l}
\dot{\rho}^2  =\dac{\Lambda}{3}\rho^2 +\dac{\lambda \kappa K}{2\rho^2}+\dac{\kappa K-2}{2}\ ,  \\ \\ \ddot{\rho} =\dac{\Lambda }{3}\rho -\dac{\lambda \kappa K}{2\rho^3}\ .
\end{array} \label{spec_a1}
\end{equation}

Let us first analyze $\Lambda=0$ case. In that case system (\ref{spec_a1}) has exact solution with integration constant which we fix from the condition $\rho\to 0$ as $t\to 0$;
the resulting solution is

\begin{equation}
\begin{array}{l}
\rho = \dac{1}{\sqrt{2}} \sqrt{t \( (\kK - 2)t + 2\sqrt{2\lambda\kK} \)}.
\end{array} \label{sol_noL}
\end{equation}

\begin{figure}
\includegraphics[width=0.4\textwidth, angle=0]{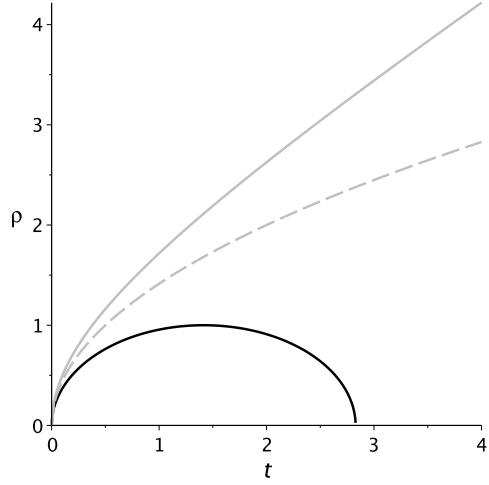}
\caption{Solutions of the $a(t)\equiv 1$ and $\Lambda = 0$ case -- $\kK < 2$ in black, $\kK = 2$ in dashed grey and $\kK > 2$ in solid grey
(see the text for more details).}\label{pre01a}
\end{figure}

One can see that for $\kK > 2$ late-time asymptote is $\rho\propto t$ while for $\kK = 2$ solution (\ref{sol_noL}) reduces to

\begin{equation}
\begin{array}{l}
\rho = \sqrt{2\sqrt{2}\lambda t},
\end{array} \label{sol_noL_kk2}
\end{equation}

\noindent and one can see that its late-time asymptote is $\rho\propto t^{1/2}$. Finally, for $\kK < 2$ the radicand in (\ref{sol_noL}) eventually
becomes negative at some $t$ which corresponds to the recollapse; all three situations are presented in Fig.~\ref{pre01a}. In black we presented $\kK < 2$ case, in dashed grey -- $\kK=2$
and in solid grey -- $\kK > 2$ cases.


\begin{figure}
\includegraphics[width=0.4\textwidth, angle=0]{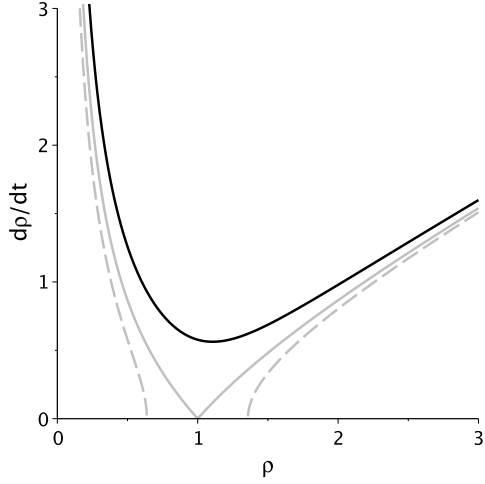}
\caption{Phase portrait of the $a(t)\equiv 1$ model with $\kK < 2$ and $\Lambda > 0$: cases with positive discriminant of (\ref{spec_a1}) (black curve), zeroth (solid grey) and negative (dashed grey)
(see the text for more details).}\label{pre01b}
\end{figure}

Now let us turn to $\Lambda\ne 0$ case. In that case we can reduce first of (\ref{spec_a1}) to biquadratic equation with respect to $\rho$ and find condition when its discriminant is negative
-- in that case $\dot\rho^2 > 0$ always. This happens if

\begin{equation}
\begin{array}{l}
\Lambda \geqslant \Lambda_0 = \dac{3}{8}~\dac{(\kK - 2)^2}{\lambda\kK}.
\end{array} \label{L1}
\end{equation}

Now let us plot $\dot\rho (\rho)$ phase portrait; we did it for $\kK < 2$ in Fig.~\ref{pre01b} for three cases -- with the discriminant of (\ref{spec_a1}) being positive (black curve),
zeroth (solid grey) and negative (dashed grey). One can see that the only smooth and nonsingular regime occurs when the discriminant is negative so if (\ref{L1}) is fulfilled. In two other
cases one faces finite-time future singularity at some finite $t$. So that to have smooth and nonsingular regime for $\kK < 2$ case we need $\Lambda > \Lambda_0$ from (\ref{L1}).
For $\kK = 2$ case, as we can see from (\ref{L1}), any $\Lambda > 0$ is sufficient; $\kK > 2$ case is unaffected by (\ref{L1}).

The late-time regime in this case is described by the $\rho(t)\to\infty$ branch from Fig.~\ref{pre01b} -- it could be derived from the first of (\ref{spec_a1}) taking the mentioned limit -- dynamical equation reduces 
to \mbox{$\dot\rho(t)^2 \simeq \Lambda\rho(t)^2/3$} with expanding solution $\rho(t) \propto \exp (\sqrt{\Lambda/3}t)$ -- usual exponential solution.

Our claim that the $\kK > 2$ case is unaffected by (\ref{L1}) could be proved as follows: from the first of (\ref{spec_a1}) one can see that for $\kK \geqslant 2$ we always have $\dot\rho^2 > 0$, 
given $\lambda,\,\Lambda,\,\kK > 0$.
Of these, $\lambda > 0$ and $K > 0$ come from the Skyrme theory and $\kappa > 0$ since we have gravitational attraction. On contrary, sometimes in different aspects of field theory $\Lambda < 0$ is
considered, which gives anti de Sitter in cosmological background. One can immediately see from the first of (\ref{spec_a1}), that in $\Lambda < 0$ case at small $\rho$ we have $\dot\rho^2 > 0$ while at
large $\rho$ it is negative, so the dynamics is limited and we have finite-time future singularity at some finite $t$, similar to the $\kK < 2$, $\Lambda < \Lambda_{cr}$ case. In case of negative $\Lambda$
it is true regardless of $\kK$, so in the remaining part of the paper we consider $\Lambda > 0$ only.

So to summarize our findings of the $a(t)\equiv 1$ case -- if $\Lambda = 0$, there are three regimes, depending on the $\kK$: if $\kK < 2$, there is a recollapse, if $\kK = 2$, the late-time regime is power-law
$\rho(t) \propto \sqrt{t}$ and if $\kK > 2$ -- it is another power-law $\rho(t) \propto t$. If $\Lambda$ is nonzero and negative, then we always have recollapse; if $\Lambda > 0$ and $\kK \geqslant 2$, we always
reach exponential regime $\rho(t) \propto \exp (\sqrt{\Lambda/3}t$. Finally, if $\Lambda > 0$ and $\kK < 2$, then if (\ref{L1}) is fulfilled, we have exponential solution and if not -- recollapse. Let us note that
all these regimes we derived analytically and so they should take place for all initial conditions; our additional numerical analysis support this claim.

\section{General case with dynamical $a(t)$}

\begin{figure}
\includegraphics[width=1.0\textwidth, angle=0]{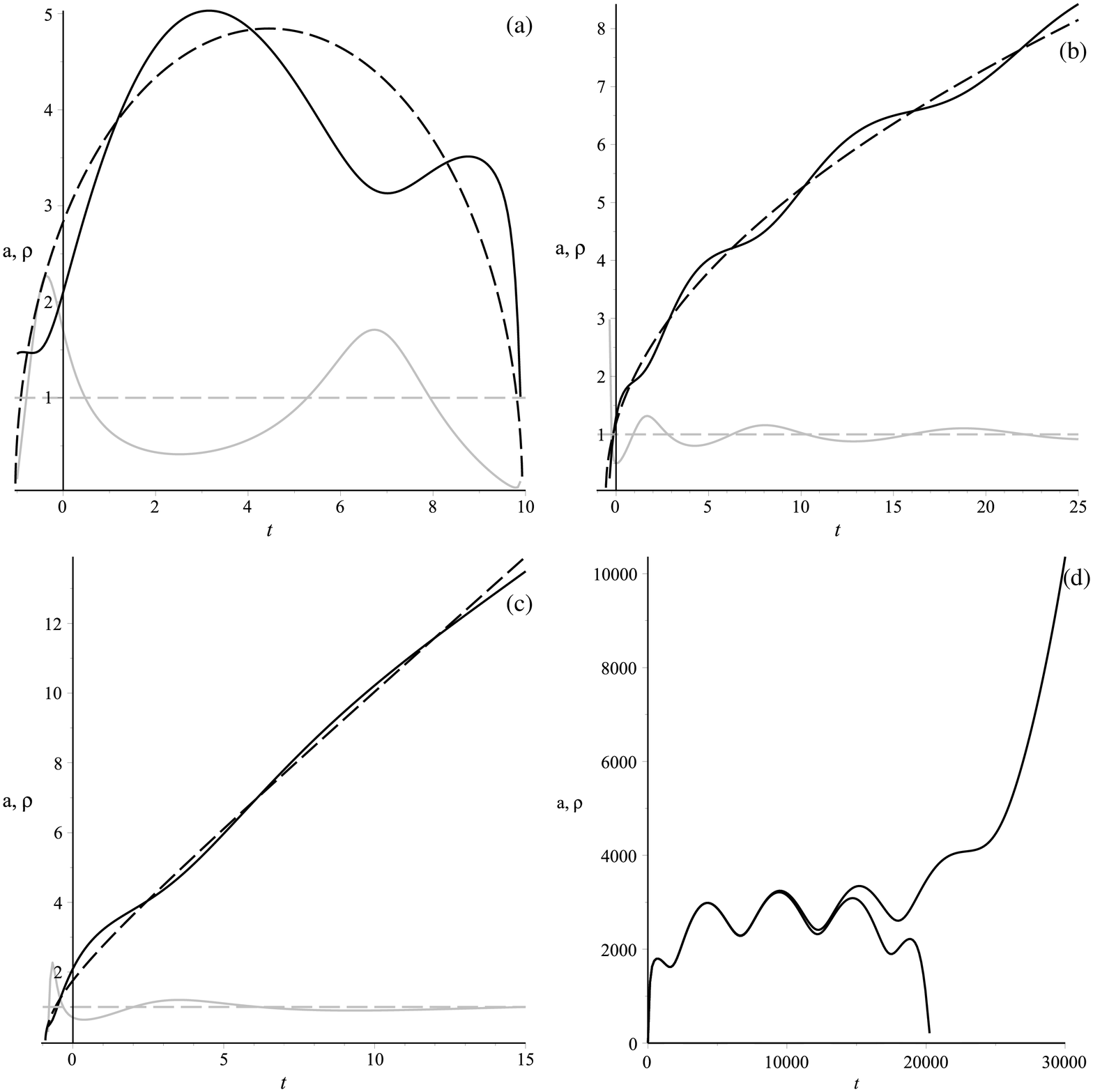}
\caption{Dynamics of $a(t) \ne 1$ and $\Lambda = 0$ case on (a)--(c) panels: $\kK < 2$ on (a) panel, $\kK = 2$ on (b) and $\kK > 2$ on (c). On (d) panel -- dynamics of $a(t) \ne 1$ and $\Lambda \ne 0$ case:
exponential (upper curve) and recollapse (lower) behavior
(see the text for more details).}\label{pre01}
\end{figure}

In this section we analyze the behavior of the general system (\ref{full}) with dynamical $a(t)$. First we numerically analyze system (\ref{full}) with $\Lambda=0$ and presented
typical behavior for each case in Fig. \ref{pre01}(a)--(c). In (a) panel we present typical behavior for $\kK < 2$ case -- one can see that it asymptotically tends to $a(t) \equiv 1$ scenario 
with oscillations around it. And
similar to the $a(t)\equiv 1$ counterpart, our dynamical $a(t)$ case has finite-time future singularity. In (b) panel we demonstrate typical $\kK=2$ dynamics -- one can see that, similar to
the previous case, we have oscillations around $a(t)\equiv 1$ regime with $\rho (t) \propto t^{1/2}$ asymptotic behavior. And finally in (c) panel we present the $\kK > 2$ case with
oscillations around $a(t)=1$ and $\rho (t) \propto t$ asymptotic behavior.
So that we can see that in all $\Lambda = 0$ cases we have oscillatory approach to the corresponding $a(t)\equiv 1$ cases, described in the previous section. Actual evolutions curves depends on the initial conditions
a bit (say, period and amplitude of oscillations depend on the initial conditions), but general behavior and late-time asymptotes are the same within the same case.

Final case to consider is general dynamical $a(t)$ with $\Lambda \ne 0$. As we just saw, with $\Lambda = 0$, dynamical $a(t)$ cases tend to their $a(t)\equiv 1$ counterparts through oscillation -- the same 
behavior have dynamical $a(t)$ cases with nonzero $\Lambda$. So, similarly to the $a(t)\equiv 1$ case, negative $\Lambda$ always leads to the recollapse regardless of $\kK$.
As we found in the previous section, $a(t)\equiv 1$ with $\Lambda > 0$ cases have either exponential regime or recollapse as a late-time attractor, and so dynamical $a(t)$ cases 
have the same attractor as well.
So for $\kK \geqslant 2$ we always have exponential solutions with damping oscillations while for $\kK < 2$ we have either recollapse or exponential solution depending on $\Lambda$ -- the same behavior we described in the
previous section for $a(t)\equiv 1$ case.
In Fig. \ref{pre01}(d) we presented typical behavior in the vicinity of separation of these two regimes -- the lower regime experience recollapses while the upper reaches exponential regime; both regimes 
experience oscillations.

\begin{figure}
\includegraphics[width=1.0\textwidth, angle=0]{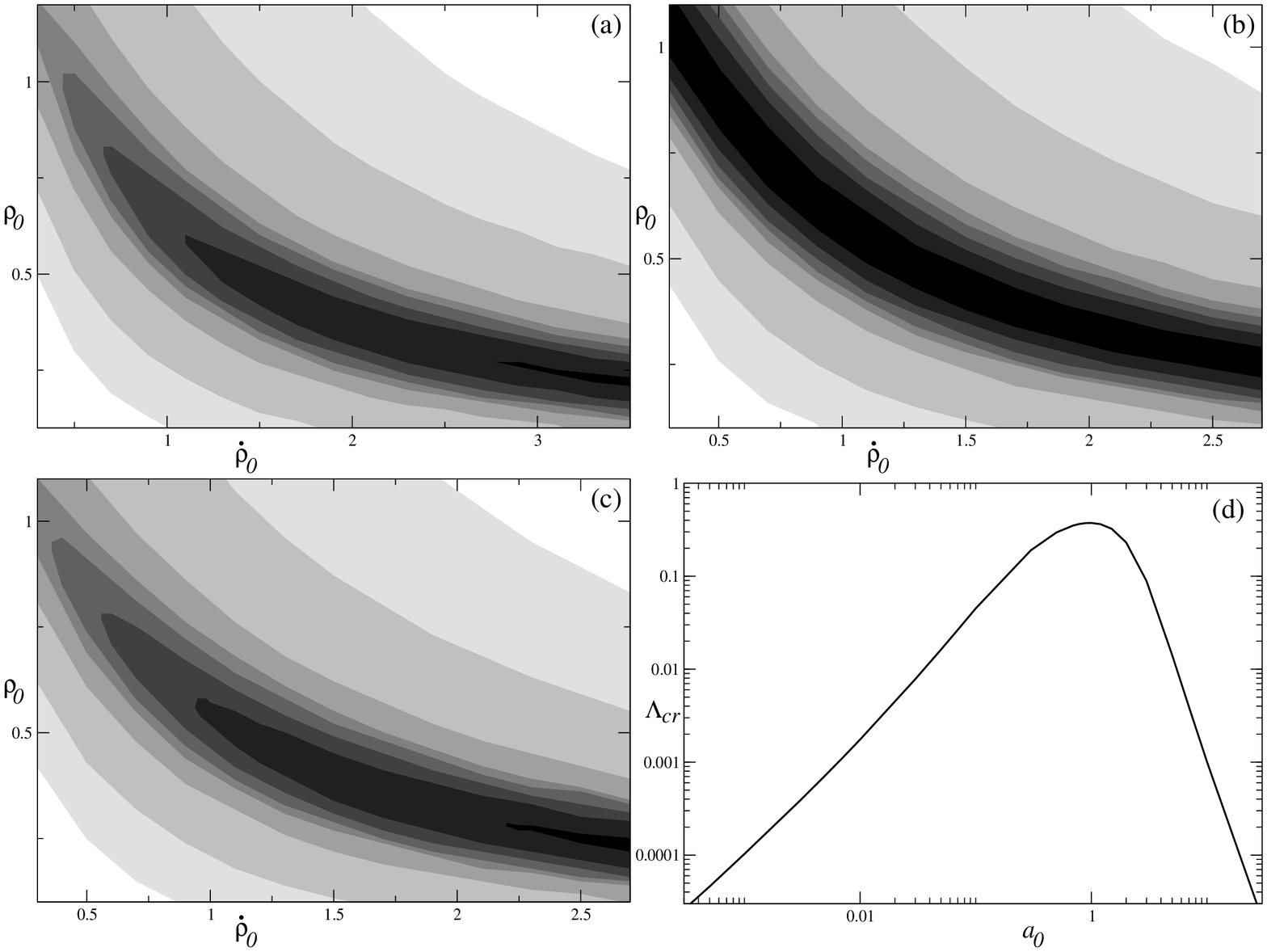}
\caption{Contours of equal $\Lambda_{cr}$ on the initial conditions space $\{\rho_0,\,\dot\rho_0\}$ for $a_0 = 0.8$ on (a) panel, $a_0 = 1.0$ on (b) panel and $a_0 = 1.2$ on (c) panel.
Example of $\Lambda_{cr}$ behavior with varying $a_0$ on (d) panel
(see the text for more details).}\label{pre02}
\end{figure}

In the general $\kK < 2$ case (with dynamical $a(t)$ with $\Lambda > 0$) the value for $\Lambda_{cr}$ which separates recollapse from exponential expansion (see this separation e.g. in
Fig. \ref{pre01}(d)) is actually lower then $\Lambda_0$, given by (\ref{L1}). Of course, generally $\Lambda_{cr} \leqslant \Lambda_0$, and actual values we present in Fig. \ref{pre02}(a)--(c).
We provided contours of equal $\Lambda_{cr}$ on the initial conditions space $\{\rho_0,\,\dot\rho_0\}$ for $a_0 = 0.8$ on (a) panel, $a_0 = 1.0$ on (b) panel and
$a_0 = 1.2$ on (c) panel.
Levels correspond to 0.37, 0.36, 0.35, 0.34, 0.33, 0.3, 0.2 and 0.1 with decreasing blackness (so black is $\Lambda_{cr} \geqslant 0.37$ and white is $\Lambda_{cr}
\leqslant 0.1$).
As these contours are plot for $\lambda=1$ and $\kK=1$ which gives $\Lambda_0 = 0.375$ derived from (\ref{L1}), we can see that for $a_0=1$, presented in Fig. \ref{pre02}(b), $\Lambda_0$ is reached for all $\rho_0$
and $\dot\rho_0$ (so that for each $\rho_0$ exists $\dot\rho_0$ where $\Lambda_0$ is reached and vice versa) -- utmost black corresponds to $\Lambda_{cr} \geqslant 0.37$. On contrary, for
$a_0$ differ from 1, $\Lambda_0$ is reached for lesser measure of the initial conditions -- see Fig.~\ref{pre02}(a) for $a_0 = 0.8$ and Fig.~\ref{pre02}(c) for $a_0 = 1.2$. We can see from these two panels
that $\Lambda_0$ is shifted towards higher $\dot\rho_0$ and with growth of $|a_0 - 1|$ difference, the gap between highest $\Lambda_{cr}$ and $\Lambda_0$ also increase -- in Fig.~\ref{pre02}(d)
we presented one-dimensional scan on $a_0$ -- one can see that $\Lambda_{cr}$ could be orders of magnitude below $\Lambda_0$.

And a short summary of this sections findings: we found that the $\Lambda = 0$ case with generic $a(t)$ have three distinct late-time regimes which coincide with those described in the previous $a(t)\equiv 1$ section.
 So for $\kK < 2$ we have a recollapse, for $\kK = 2$, the late-time regime is power-law $\rho(t) \propto \sqrt{t}$ and for
$\kK > 2$ -- it is another power-law $\rho(t) \propto t$. In the general $\Lambda > 0$, dynamical $a(t)$ case, again, similar to the $a(t)\equiv 1$ case, we have either exponential solution or recollapse. The former
of them takes place for $\kK \geqslant 2$ while the latter for $\kK < 2$ and $\Lambda < \Lambda_{cr}$. This $\Lambda_{cr} \leqslant \Lambda_0$ defined from (\ref{L1}) and the actual value for $\Lambda_{cr}$ heavily depends 
on the initial conditions, as presented in Fig.~\ref{pre02}. One cannot miss the strong dependence of $\Lambda_{cr}$ on $a_0$ -- more initial anisotropy -- lesser value for $\Lambda$-term is needed to reach 
exponential expansion.

\section{Linear stability}

Now let us turn our attention to the stability of the solutions. In the course of paper we saw there are three nonsingular regimes: two power-law -- $\rho(t) \propto \sqrt{t}$ and
$\rho(t) \propto t$, and exponential $\rho(t) \propto \exp(Ht)$; all three regimes have $a(t) \to 1$. So that
we perturb full system (\ref{full}) around solution $a(t) = 1$ and with these three different $\rho(t)$. Linear perturbations around $a(t) = 1$ read
$a\to 1+\delta a$, $\dot a \to \dot\delta a$, $\ddot a \to \ddot\delta a$, $\rho\to\rho+\delta\rho$, $\dot\rho\to\dot\rho+\dot\delta\rho$,
$\ddot\rho\to\ddot\rho+\ddot\delta\rho$ and the equations on perturbations take form

\begin{equation}
\begin{array}{l}
4\rho^3\dot\rho\dot{\delta a} + 12\rho^2\dot\rho\dot{\delta\rho} + (-4\Lambda\rho^4 + 12\rho^2\dot\rho - 4\rho^2\kK + 8\rho^2 - 2\lambda\kK) \delta a + \\ + (-8\Lambda\rho^3 + 12\rho\dot\rho - 6\rho\kK + 12\rho)\delta\rho = 0, \\ \\
4\rho^3 \ddot{\delta\rho} + 4\rho^2\dot\rho\dot{\delta\rho} + (-4\Lambda\rho^4 + 8\rho^3\ddot\rho + 4\rho^2\dot\rho^2 - 4\rho^2\kK - 8\rho^2 - 2\lambda\kK)\delta a + \\ + (-8\Lambda\rho^3 + 12\rho^2\ddot\rho
+ 4\rho\dot\rho^2 - 2\rho\kK + 4\rho)\delta\rho = 0, \\ \\
\rho^4 \ddot{\delta a} + 3\rho^3\dot\rho\dot{\delta a} + (2\rho^2 \kK + 8\rho^2 + 2\lambda\kK) \delta a = 0.
\end{array} \label{pert}
\end{equation}

Last of (\ref{pert}) could be solved for stability in $a$-direction. Substitution of exponential solution $\rho(t) = \rho_0 \exp (Ht)$ leads us to

\begin{equation}
\begin{array}{l}
\rho_0^4 \exp (4Ht) \( \ddot\delta a(t) + 3H\dot \delta a(t) \) + 2\rho_0^2 \exp (2Ht) \delta a(t) (\kK + 4) + 2\kK \lambda \delta a(t) = 0.
\end{array} \label{pert_exp_1}
\end{equation}

\noindent Using new variable $y = \dot\delta a(t)/\delta a(t)$, we can rewrite (\ref{pert_exp_1}) as

\begin{equation}
\begin{array}{l}
\dot y + y^2 + 3Hy + F(t) = 0,~~F(t) = \dac{2\kK}{\rho_0^2 e^{2Ht}} + \dac{8}{\rho_0^2 e^{2Ht}} + \dac{2\lambda\kK}{\rho_0^4 e^{4Ht}},
\end{array} \label{pert_exp_2}
\end{equation}

\noindent where $F(t)$ could be treated as a perturbative force acting on a system described by homogeneous equation. The solution of the homogeneous equation from (\ref{pert_exp_2}) is

\begin{equation}
\begin{array}{l}
y(t) = \dac{3H}{3HC_1 e^{3Ht} - 1},
\end{array} \label{pert_exp_3}
\end{equation}

\noindent and then we can solve it for $\delta a(t)$:

\begin{equation}
\begin{array}{l}
\delta a(t) = C_2 \( 3HC_1 - e^{-3Ht}\).
\end{array} \label{pert_exp_4}
\end{equation}

The solution of the general (\ref{pert_exp_1}) equation leads to an expression through M and W Whittaker functions~\cite{spec_func} and generally cannot be expressed through elementary functions. 
But with an analysis performed in (\ref{pert_exp_2})--(\ref{pert_exp_4}) we describe the general behavior as follows: $F(t)$ acts as a perturbative force and generates oscillations around (\ref{pert_exp_3})
-- the solution of the homogeneous equation from (\ref{pert_exp_2}). One can see that at $t\to\infty$ we have $F(t)\to 0$ so that at late times we can use (\ref{pert_exp_3}) as an exact solution, which
leads to (\ref{pert_exp_4}) as a solution for original perturbation equation (\ref{pert_exp_1}). One can note that the amplitude of perturbations does not damp to zero -- as $t\to\infty$ we have
$\delta a \to 3HC_1 C_2$. The reason behind it is not clear, but as the perturbations do not grow, we treat this case as stable.
Our numerical
analysis totally supports this description -- 
at the beginning the solution is represented by damping oscillations, but after they decay the asymptote value is not zero but some small constant. This is the same for a wide variety 
of the initial conditions and the constant is also the same; though it varies for different parameters. 

Now let us turn our attention to the power-law regimes. In that case the solution of the last of (\ref{pert}) could be written in terms of J and Y Bessel functions and is represented by oscillations with
damping amplitude, which directly points to stability, as long as solution itself exists. Solution for $\rho(t) = \rho_0 \sqrt{t/t_0}$ exists iff $\rho_0^4 \geqslant 64\lambda t_0^2$ and solution for
$\rho(t) = \rho_0 (t/t_0)$ exists iff $\rho_0^2 \geqslant 2(\kK + 4)$.

To summarize, we found that the exponential solution behave a bit unusually, but we claim that we could call it stable -- the perturbations experience damping oscillations and reach constant value afterwards.
As they do not grow, we claim them to be stable. The power-law solutions are stable everywhere within their range of existence.

\section{Discussion}

In current paper we considered Bianchi-IX cosmological model in Einstein-Skyrme system (\ref{full}). The original system was simplified and considered with growth of complexity, which allows us to build
semi-analytical solution. Purely analytical solutions are obtained for the simplest case with $a(t)\equiv 1$ and $\Lambda=0$ -- in that case there are three possible solutions -- one with recollapse for
$\kK<2$ and two power-laws -- $\propto t^{1/2}$ for $\kK=2$ and $\propto t$ for $\kK > 2$. All three are presented in Fig. \ref{pre01a} and one cannot miss their similarity with three different
Friedmann solutions from
classical cosmology -- with spatial curvature $k=\pm 1$ and $0$. The scales with time are different but the qualitative behavior is the same -- in some sense $(2-\kK)$ plays a role similar to the spatial
curvature.

Further complications of the system act as modifications of the obtained exact solution. Turning $a(t)$ dynamical (but with still $\Lambda=0$) leads to oscillatory behavior like presented in
Fig.~\ref{pre01}(a)--(c). Let us remind that oscillatory behavior is a part of early Bianchi-IX universe, as discovered by Belinskij, Khalatnikov and Lifshits~\cite{belinski}. If one keep $a(t)\equiv 1$ but
make
$\Lambda > 0$, then power-law regimes turn to exponential while recollapse regime turn to exponential if (\ref{L1}) is satisfied; if not, they remain recollapse. Finally, if one combine both -- dynamical
$a(t)$
with $\Lambda > 0$, the resulting trajectories have oscillations and exponential (de Sitter) late-time asymptote for $\kK \geqslant 2$; for $\kK < 2$ one have oscillations and de Sitter if
$\Lambda > \Lambda_{cr}$
and recollapse if $\Lambda < \Lambda_{cr}$; the separation between these two cases is presented in Fig. \ref{pre01}(d).
Recollapse behavior is also encountered in anti de Sitter case -- when $\Lambda < 0$ -- and in this case the result is independent on $\kK$.
The value for $\Lambda_{cr}$ cannot exceed $\Lambda_0$ from (\ref{L1}) but could be much less (orders of magnitude), as our numerical investigation suggests. In
Fig.~\ref{pre02} we provided the distribution of $\Lambda_{cr}$ over initial conditions space for three different $a_0$ on (a)--(c) panels and linear cut over $a_0$ on (d).

One can see that all nonsingular regimes have $a(t) \to 1$ at late times. From metric (\ref{BIX}) point of view, $a(t) = 1$ solution is the most symmetric one (so that it has more Killing fields then
$a(t) \ne 1$ one), so that we can see that all nonsingular regimes have symmetry restoring dynamics, and all these solutions are stable. Singular regimes, which do not possess this feature, are either
$\kK < 2$ cases with $\Lambda < \Lambda_{cr}$ or $\Lambda < 0$ AdS cases; for the latter the value for $\kK$ is irrelevant.

For more physical analysis we use real values for the Skyrme coupling constants~\cite{real}. Then one can immediately see that $\kK \lll 1$ and so $\kK < 2$ is the case. For $\kK < 2$ from (\ref{sol_noL})
one can derive the ``lifetime'' -- with real values for couplings substituted, this time appear to be of the order of Planck time, which means that without $\Lambda$-term or some other matter sources with
sufficient density, Bianchi-IX universe with Skyrme would collapse immediately. On the other hand, on this time scales the space-time cannot be described by classical means and additional investigation
with involvement of quantum physics is required.
Finally, if we substitute coupling constants into (\ref{L1}), the resulting value for the cosmological constant appears to be
in agreement with other estimates from quantum field theory, treating it as vacuum energy, and is around 120 orders of magnitude higher than the observed value (so-called ``cosmological constant problem'',
see e.g.~\cite{weinberg}).

In a sense the results of current paper complement the results of~\cite{we14}, where we studied Bianchi-I and Kantowski-Sachs universes in Einstein-Skyrme system. In both papers the cosmological constant
(or probably some other matter field) is necessary for viable cosmological behavior. But unlike~\cite{we14}, where we demonstrated need for the upper bound on the value of $\Lambda$-term, in current paper
we found the lower bound. It is interesting that different topologies in presence of Skyrme source require either not too large or not too low values for the cosmological constant.

This finalize our study of Bianchi-IX Skyrme-Einstein system. We described its dynamics and derived conditions for different regimes to take place.
Generally, Einstein-Skyrme
systems are very interesting and are not much considered, probably due to their complexity, so each new result improves our understanding of cosmological hadron dynamics. In particular, these systems
offer the interesting possibility to study the cosmological consequences to have conserved topological charge. Thus the present analysis is quite relevant as the energy-momentum tensor a Skyrmions of
unit topological charge.

\begin{acknowledgments}
This work was supported by FAPEMA under project BPV-00038/16. We are thankful to the referee for their valuable comments which lead to the manuscript improvement.
\end{acknowledgments}

\end{document}